\documentclass[preprint,final,5p,times,twocolumn]{elsarticle}

\usepackage[colorlinks,linkcolor=blue,anchorcolor=blue,citecolor=blue,urlcolor=blue]{hyperref}
\usepackage{graphicx}
\usepackage{epstopdf}
\usepackage{dcolumn}
\usepackage{textcomp}
\usepackage{gensymb}
\usepackage{booktabs}
\usepackage{amsmath}
\usepackage{amssymb}
\usepackage{color}
\usepackage{hyperref}
\usepackage{url}
\usepackage{siunitx}
\usepackage{subfigure}
\usepackage{longtable}
\usepackage{array}
\usepackage{booktabs}
\usepackage{verbatim}
\usepackage{threeparttable}
\usepackage{orcidlink}

\emergencystretch=\maxdimen
\usepackage{dcolumn}
\newcolumntype{x}{D{.}{.}{6.6}}
\newcolumntype{y}{D{.}{.}{4.5}}
\newcolumntype{z}{D{.}{.}{5.7}}
\newcolumntype{f}{D{.}{.}{7.9}}
\newcolumntype{e}{D{.}{.}{5.6}}

\journal{Physics Letters B}

\begin{document}
\begin{frontmatter}

\title{Nuclear charge radii of germanium isotopes around $N$ = 40}

\author[PKU]{S.~J.~Wang}
\author[KULEUVEN]{A.~Kanellakopoulos\fnref{fn1}}
\author[PKU,KULEUVEN]{X.F. ~Yang\corref{cor1}}\ead{xiaofei.yang@pku.edu.cn}
\author[PKU]{S.~W.~Bai}
\author[TUM]{J.~Billowes}
\author[TUM]{M.~L.~Bissell}
\author[MPI]{K. ~Blaum}
\author[LIVEPOOL]{\mbox{B. ~Cheal}}
\author[LIVEPOOL]{C.~S.~Devlin}
\author[MIT,CERN]{\mbox{R.~F.~Garcia Ruiz}}
\author[THU]{J.~Z.~Han}
\author[CERN]{H.~Heylen}
\author[TUD,MAINZ]{S.~Kaufmann}
\author[TUD]{K.~K\"{o}nig}
\author[KULEUVEN]{\mbox{\'{A}.~Koszor\'{u}s}}
\author[CERN,TU]{S.~Lechner}
\author[CERN]{S.~Malbrunot-Ettenauer\fnref{fn2}}
\author[MSU]{W.~Nazarewicz}
\author[MPI,MAINZ]{R.~Neugart}
\author[KULEUVEN,CERN]{G.~Neyens}
\author[TUD]{W.~N\"{o}rtersh\"{a}user}
\author[TUD]{\mbox{T.~Ratajczyk}}
\author[PG]{P.-G.~Reinhard}
\author[MPI,CERN]{L.~V.~Rodr\'{i}guez}
\author[KULEUVEN,CERN]{S.~Sels}
\author[TUM]{L.~Xie}
\author[KULEUVEN]{Z.~Y.~Xu\fnref{fn3}}
\author[IPN]{D.~T.~Yordanov}
\author[CAS,UCAS]{Y.~M.~Yu}

\cortext[cor1]{Corresponding author}
\fntext[fn1]{Present address: HEPIA, HES-SO University of Applied Scienes and Arts Western Switzerland, Geneva, Switzerland}
\fntext[fn2]{Present address: TRIUMF, 4004 Wesbrook Mall, Vancouver, BC V6T 2A3, Canada}
\fntext[fn3]{Present address: Department of Physics and Astronomy, University of Tennessee, 37996 Knoxville, TN, USA}

\address[PKU]{School of Physics and State Key Laboratory of Nuclear Physics and Technology, Peking University, Beijing 100871, China}

\address[KULEUVEN]{KU Leuven, Instituut voor Kern- en Stralingsfysica, B-3001 Leuven, Belgium}

\address[TUM]{School of Physics and Astronomy, The University of Manchester, Manchester M13 9PL, United Kingdom}

\address[MPI]{Max-Planck-Institut f\"{u}r Kernphysik, D-69117 Heidelberg, Germany}

\address[LIVEPOOL]{Oliver Lodge Laboratory, Oxford Street, University of Liverpool, Liverpool, L69 7ZE, United Kingdom}

\address[MIT]{Massachusetts Institute of Technology, Cambridge, MA, USA}
\address[CERN]{Experimental Physics Department, CERN, CH-1211 Geneva 23, Switzerland}

\address[THU]{State Key Laboratory of Precision Measurement Technology and Instruments, Key Laboratory of Photon Measurement and Control Technology of Ministry of Education, Department of Precision Instrument, Tsinghua University, Beijing 100084, China}

\address[TUD]{Institut f\"{u}r Kernphysik, TU Darmstadt, D-64289 Darmstadt, Germany}
\address[MAINZ]{Institut f\"{u}r Kernchemie, Universit\"{a}t Mainz, D-55128 Mainz, Germany}

\address[TU]{Technische Universit\"{a}t Wien, Karlsplatz 13, AT-1040 Wien, Austria}

\address[MSU]{Department of Physics and Astronomy and FRIB Laboratory, Michigan State University, East Lansing, Michigan 48824, USA}

\address[PG]{Institut f{\"u}r Theoretische Physik, Universit{\"a}t Erlangen, Erlangen, Germany}

\address[IPN]{Institute de Physique Nucl\'eaire, CNRS-IN2P3, Universit\'e Paris-Sud,Universit\'e Paris-Saclay, 91406 Orsay, France}

\address[CAS]{Beijing National Laboratory for Condensed Matter Physics, Institute of Physics, Chinese Academy of Sciences, Beijing 100190, China}
\address[UCAS]{University of Chinese Academy of Sciences, Beijing 100049, China}

\begin{abstract}
Collinear laser spectroscopy measurements were performed on $^{68-74}$Ge isotopes ($Z = 32$) at ISOLDE-CERN, by probing the $4s^2 4p^2 \, ^3\!P_1 \rightarrow 4s^2 4p 5s \, ^3\!P_1^o$ atomic transition (269~nm) of germanium. Nuclear charge radii are determined via the measured isotope shifts, revealing a larger local variation than the neighboring isotopic chains. Nuclear density functional theory with the Fayans functionals Fy($\Delta r$,HFB) and Fy(IVP), and the SV-min Skyrme describes the experimental data for the differential charge radii $\delta\langle r^{2} \rangle$ and charge radii $R_{\rm c}$ within the theoretical uncertainties. The observed large variation in the charge radii of germanium isotopes is better accounted for by theoretical models incorporating ground state quadrupole correlations. This suggests that the polarization effects due to pairing and deformation contribute to the observed large odd-even staggering in the charge radii of the Ge isotopic chain.
\end{abstract}


\end{frontmatter}

\section{Introduction}

The charge radius is one of the basic properties of atomic nuclei. Based on general considerations, the nuclear charge radii are expected to globally scale with the nuclear mass number $A$ as $A^{1/3}$. It is known, however, that the local deviations from this bulk behavior are large; they are due to various effects, including  shell structure variations, shape deformations, nuclear superconductivity, and continuum coupling~\cite{PPNP2023}. This sensitivity makes nuclear charge radii  excellent laboratories of structural phenomena such as halo structures~\cite{Be-halo,Li-halo,He-halo}, nuclear magicity and nucleonic pairing~\cite{Ca-radii2016,K-radii2021,Sc-radii,Ni-radii,Sn-radii}, nuclear correlations~\cite{Cu-radii,Zn-radii2019,Cu-radii2020}, and static shape deformations and shape coexistence~\cite{Zn-radii2016,Hg-radii2018,Bi-radii-2021,Au-radii}. The nuclear charge radius has thus become one of the key observables to explore the structure of atomic nuclei in different regions of the chart of nuclides~\cite{Be-halo,K-radii2021,Sn-radii,Ni-radii2022,No-radii} and to provide benchmarks for the tests of nuclear many-body methods and interactions~\cite{Fayan2017,NNLOgo,Kortelainen2022}.

The region of nuclei around the semi-magic $^{68}$Ni and doubly magic $^{78}$Ni
has been a topic of intensive experimental and theoretical studies~\cite{Sorlin2002, Langanke2003,Heyde2011,78NiPRL, 78Ninature,74Zn-PRL,78Ni-Sc-Theo,Cu-79-mass,Cu-79}. In the last decade, a series of experiments has been conducted in order to gain a deeper understanding of the nuclear structure in this region. These experimental investigations include high-precision laser spectroscopy and mass measurements, resulting in rich data on the nuclear properties, such as nuclear masses~\cite{Cu-79-mass,Zn-79-mass,Ni-region-mass}, spins~\cite{Zn-radii2016, Flanagan2009}, electromagnetic moments~\cite{Wraith2017,Cheal2010,Ge-moments}, and charge radii~\cite{Ni-radii,Zn-radii2019,Cu-radii2020}. Insightful information on shell evolution~\cite{Flanagan2009,Wraith2017,Cheal2010}, emergence of collectivity around $N = 40$~\cite{Zn-radii2019,Ni-region-mass,Cheal2010,Yang2018}, and shape coexistence~\cite{Zn-radii2016,Zn-79-mass} near $^{78}$Ni has been obtained from these measurements.

More specifically, based on earlier experimental measurements of the nuclear charge radii in this mass region including nickel ($Z = 28$)~\cite{Ni-radii2022}, copper ($Z =29$)~\cite{Cu-radii,Cu-radii2020}, zinc ($Z = 30$)~\cite{Zn-radii2019}, and gallium ($Z = 31$)~\cite{Ga-radii-2012,Ga-radii-2017}, a change of nuclear structure has been observed around $N = 40$ with protons being added above the $Z=28$ shell closure~\cite{PPNP2023}. The ground states of germanium and selenium isotopes around $N = 40$  are  expected to have a fairly complex structure due to large quadrupole correlations and shape coexistence effects~\cite{Heyde2011} resulting in configuration mixing. According to the
multiple Coulomb excitation analysis~\cite{Sugawara2003}, the ground state configurations of $^{70,72,74,76}$Ge are associated with modest deformations. The excited $0^+_2$ state is well deformed in  $^{70}$Ge and it is nearly spherical in $^{74,76}$Ge. In the nucleus $^{72}$Ge, the two
lowest 0$^+$ states are strongly mixed~\cite{Sugawara2003,Ayangeakaa2016}, as evidenced by their similar quadrupole invariants $\langle Q^2 \rangle$ \cite{Kumar1972,Cline1986}~(For the shape mixing analysis of   $^{70,72,74,76}$Ge, see Fig. 37 of Ref.~\cite{Heyde2011} and Ref.~\cite{Gupta2019}.).

Theoretical potential energy surfaces for the Ge isotopic chain~\cite{Guo2007,Shen2011,Corsi2013,Niksic2014,Nomura2017,Nomura2022} are predicted to be fairly soft, with shallow coexisting spherical, prolate, and oblate minima. In some cases, $\gamma$-soft or triaxial shapes are expected~(For a comprehensive analysis of quadrupole invariants in the Ge chain, see Ref.~\cite{Ayangeakaa2016}.).  Due to the transitional character of the nuclei considered, there is a considerable spread between predictions of different theoretical models.
The question that will be addressed in this paper is: ``In which way does the complex pattern of quadrupole collectivity of the Ge isotopes impact their charge radii?''

Prior to the present measurements, laser spectroscopic information on the charge radii of germanium isotopes, even for the stable ones $^{70,72-74,76}$Ge, is scarce~\cite{radiibook,radiireview}. This is due to the numerous experimental challenges, such as the difficulty to produce radioactive germanium in ionic form at ISOL facilities, and the optical transition which is not easily accessible by commonly used laser devices, as detailed in Ref.~\cite{Ge-moments}. Here, we report the first nuclear charge radii measurement of radioactive germanium isotopes using high-resolution collinear laser spectroscopy, benefiting from the combination of laser frequency-mixing and second harmonic generation to produce 269-nm light ~\cite{Ge-moments}. The measured root mean squared (rms) charge radii of $^{68-74}$Ge isotopes are compared with the adjacent isotopic chains revealing an enhanced odd-even staggering (OES) effect around $N=40$. Our results are interpreted on the basis of nuclear density functional theory (DFT) using three energy density functionals, highlighting the impact of pairing, deformation, and zero-point ground-state quadrupole correlations (GSC) on nuclear charge radii.

\begin{figure}[t!]
\centering
\includegraphics[width=0.49\textwidth]{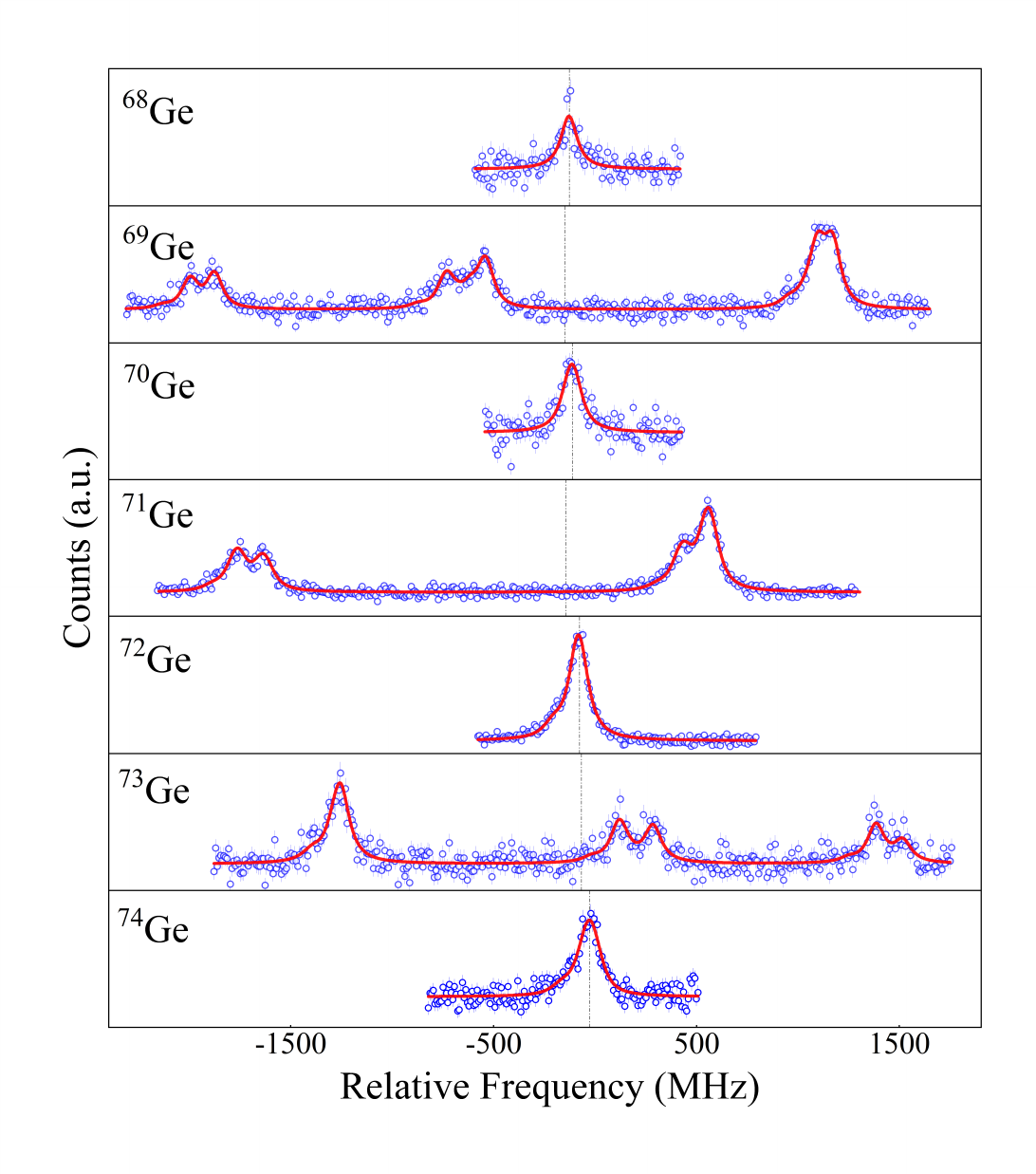}
\caption{Hyperfine structure spectra of $^{68-74}$Ge isotopes measured at COLLAPS for the atomic germanium transition of $4s^2 4p^2 \, ^3\!P_1 - 4s^2 4p 5s \, ^3\!P_1^o$. The red solid lines and vertical dashed lines show the fits to the data and the center of gravity of each spectrum, respectively. Stable isotope $^{72}$Ge was used as the reference for the isotope shifts measurements to $^{68-74}$Ge.}
\label{fig1}
\end{figure}

\section{Experimental method}
\label{sec2}
The experiment was performed at the COLLAPS setup~\cite{Neugart2017} at ISOLDE-CERN. Details of the general setup and experiment can be found in Ref.~\cite{Ge-moments}. In brief, the germanium isotopes were produced by impinging 1.4-GeV protons on a $\textrm{ZrO}_2$ target and the released atoms were ionized by a plasma ion source. The extracted germanium ions were accelerated to 50\;keV and then mass separated. Prior to being delivered to the COLLAPS beamline, the ions were first cooled in a gas-filled linear Paul trap (ISCOOL)~\cite{Mane2009} for 5~ms and released as bunches with a temporal width of 5 $\mu$s. At COLLAPS, the germanium ions were neutralized in-flight by passing through a charge-exchange cell (CEC) filled with sodium vapor and the ground state $4s^2 4p^2 \, {^3\!}P_1$ of Ge I was populated~\cite{Vernon2019}. The atomic germanium beam was then collinearly overlapped with a frequency-mixed and frequency-doubled continuous-wave (cw) laser beam at a wavelength of 268.9~nm to match the Doppler shifted atomic transition $4s^2 4p^2 \, {^3\!}P_1 - 4s^2 4p 5s \, {^3\!}P_1^o$. The laser frequency was stabilized by a wavelength meter, which was regularly calibrated with a stabilized diode laser locked to one of the hyperfine components of $^{87}$Rb atom. The kinetic energy of the germanium beam was tuned by applying a varying voltage to the CEC, in order to resonantly excite the germanium atoms into its $4s^2 4p 5s \, {^3\!}P_1^o$ state through Doppler tuning~\cite{PPNP2023}. Four photomultiplier tubes were used to detect the subsequently emitted fluorescence photons as a function of the tuning voltage~\cite{Ge-moments}. In this manner, hyperfine structure (hfs) spectra of germanium isotopes were recorded. To compensate for any possible long-term drifts in the ion velocity or in the laser frequency, hfs spectra of all studied germanium isotopes were measured alternating with the stable reference isotope $^{72}$Ge.

\section{Experimental results}
Typical hfs spectra of $^{68-74}$Ge isotopes shown in Fig.~\ref{fig1} were analyzed using a $\chi^2$-minimization approach implemented in the SATLAS package~\cite{Gins2018}. The extracted magnetic and quadrupole hfs parameters $A$ and $B$ of the odd-$A$ isotopes have already been reported in Ref.~\cite{Ge-moments}. Based on the extracted center of gravity (COG) of each spectrum,  $\nu^{A}$, isotope shifts $\delta\nu^{72,A}=\nu^{A}-\nu^{72}$ of $^{68-74}$Ge isotopes were calculated with respect to the COG ($\nu^{72}$) of the reference isotope $^{72}$Ge, as presented in Table~\ref{tab1}. The differential mean-square charge radii $\delta\langle r^{2} \rangle^{72,A}\equiv\langle r^{2} \rangle^{A}-\langle r^{2} \rangle^{72}$ can then be calculated from the isotopes shifts ($\delta\nu^{72,A}$) according to:
\begin{equation}
\delta \nu ^{72,A} = K_{\rm MS}\frac{m_{A}-m_{72}}{m_{A}m_{72}}+F \delta \langle r^{2}\rangle^{72,A}
\end{equation}
where $K_{\rm {MS}}$ and $F$ are the atomic mass-shift and field-shift factors, respectively, of the probed atomic transition, and $m_{A}$ are the atomic masses taken from Ref.~\cite{AME2021}.

\begin{table}[!t]
\caption{Isotope shifts $\delta \nu^{72,A}$, changes in mean-square charge radii $\delta\langle r^{2}\rangle^{72,A}$ and absolute charge radii $R_{\rm ch}$  of $^{68-74}$Ge isotopes. Statistical errors are shown in curved brackets. Systematic errors in square brackets arise from the uncertainty on the atomic mass-shift and field-shift factors. The experimentally calibrated atomic factors of $F = +317(58)$~MHz/fm$^{2}$ and $K_{\rm{MS}}=-9(22)$~GHz~u are used for the calculation of the charge radii. See text for more details.}
\vspace{2mm}
\renewcommand*{\arraystretch}{1.2}
\begin{tabular}{l|lllll}
 $A$& $I^{\pi}$ &$T_{1/2}$ &$\delta \nu^{72,A}$ &$\delta\langle r^{2}\rangle^{72,A}$ & $R_{\rm ch}$ \\
  &            &           & (MHz)     &  (fm$^2$)        &   (fm)           \\
\hline
68& 0$^{+}$    &  270.9 d  & -51(4)    &  -0.184(13)[67]   &     4.036(3)[8]  \\
69& 5/2$^{-}$  &  39.05 h  & -72(2)    &  -0.244(6)[62]    &     4.029(2)[8]  \\
70& 0$^{+}$    & Stable    & -35(2)    &  -0.121(6)[36]    &     4.044(2)[4]  \\
71& 1/2$^{-}$  & 11.43 d   &  -69(3)   &  -0.223(9)[43]    &     4.032(2)[5]  \\
72& 0$^{-}$    & Stable    & 0         &  0             &     4.059(2)[0]$^{\rm a}$\\
73& 9/2$^{+}$  & Stable    & 8(4)      &  0.031(13)[15]    &     4.063(3)[2] \\
74& 0$^{-}$    & Stable   & 52(2)     &  0.174(6)[41]     &     4.080(2)[5]  \\
\end{tabular}
\label{tab1}
\begin{tablenotes}
\item[a] a: $\langle r^{2}\rangle_{\rm{\mu e}}^{1/2}$ calculated from the electron scattering and muonic atom spectroscopy data~\cite{radiibook}.
\end{tablenotes}
\end{table}

In general, the atomic mass-shift and field-shift factors can be calibrated from the experimentally known charge radii of at least three stable isotopes using a King-plot analysis~\cite{radiibook} or calculated by using advanced atomic theory~\cite{PPNP2023}. For germanium isotopes, nuclear radius parameters $\varLambda_{\rm{\mu e}}^{72,70}$, $\varLambda_{\rm{\mu e}}^{74,72}$, which in a first approximation is equal to differences in the mean square charge radii $\delta\langle r^{2}\rangle$, are known experimentally from a combined analysis of electron scattering and muonic atom spectroscopy data~\cite{radiibook}. Therefore, a King plot analysis~\cite{radiibook} has been performed using the measured $\delta\nu^{72,A}$, leading to the atomic factors of $F$ = +317(58)~MHz/fm$^{2}$ and $K_{\rm{MS}}=-9(22)$~GHz~u. To further validate these calibrated atomic factors, we performed atomic calculations using the configuration interaction approach with many-body perturbation theory (CI$+$MBPT) implemented in the AMBiT software~\cite{KAHL2019232}, resulting in $F=453(100)$~MHz/fm$^{2}$ and $K_{\rm{MS}}=-14(5)$~GHz~u, where theoretical uncertainties are estimated by considering the possible impact due to the higher-order electronic correlations. Within the uncertainty, these calculated atomic factors are in agreement with those obtained from the King-plot analysis.

\begin{figure}[t!]
\centering
\includegraphics[width=0.49\textwidth]{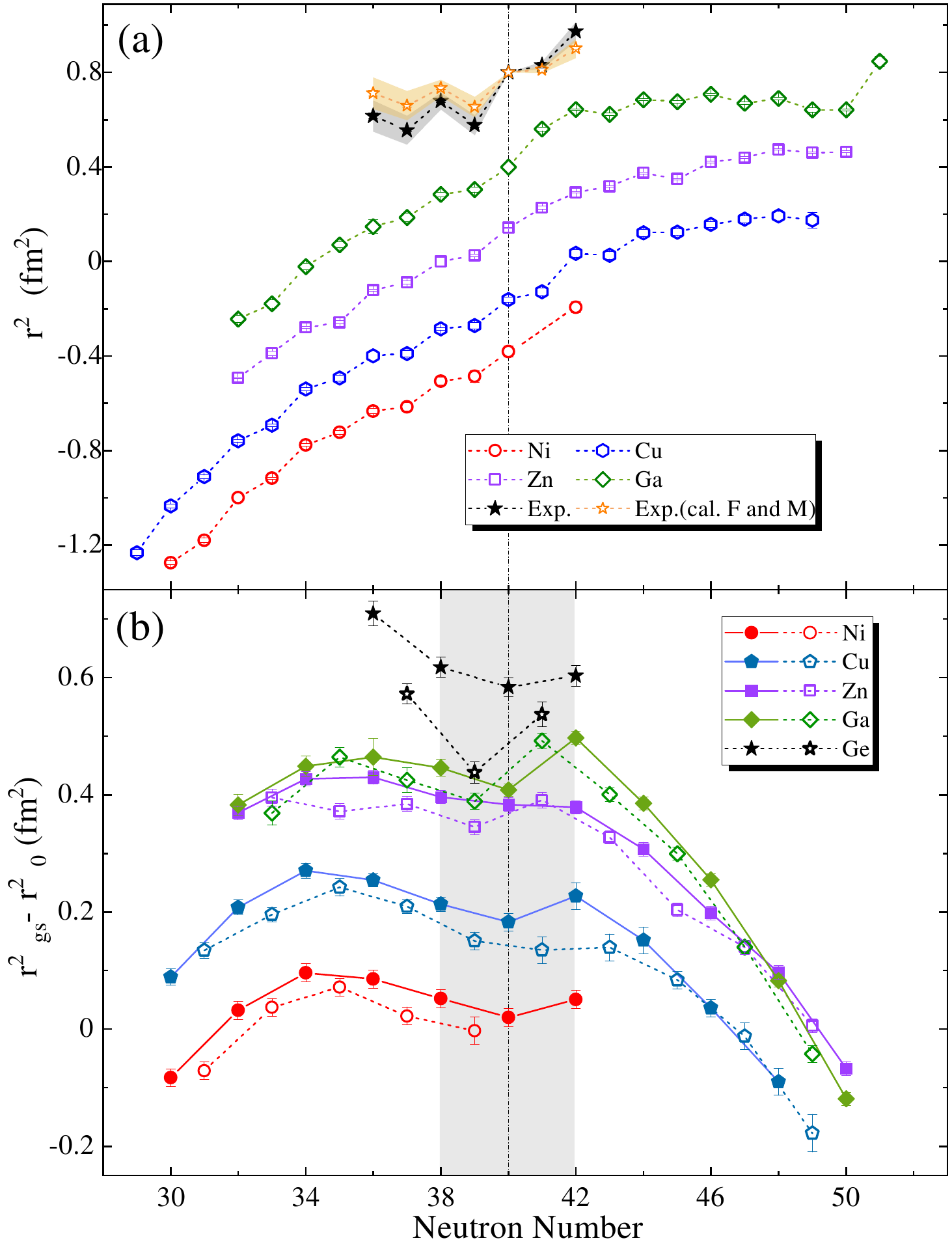}
\caption{(a) Changes in mean-square charge radii of the $^{68-74}$Ge isotopes compared with the neighbouring isotopic chains of nickel, copper, zinc and gallium. Note that the data are arbitrarily offset for better visualization. The brown empty stars present the radii extracted using the $F$ and $M$ factors from CI+MBPT calculations. (b) $\langle r^{2} \rangle_{\rm{gs}}-\langle r^{2} \rangle_{0}$ of ground state of the nickel, copper, zinc, gallium and germanium isotopes obtained by subtracting the spherical droplet model reference $\langle r^{2} \rangle_{0}$~\cite{Berdi1985}. See text for more details.}
\label{fig2}
\end{figure}

Table~\ref{tab1} and Fig.~\ref{fig2}(a)~(solid black stars) show
the differential charge radii $\delta\langle r^{2} \rangle^{72,A}$ of $^{68-74}$Ge calculated by using the calibrated atomic factors. The systematic uncertainties for $\delta\langle r^{2} \rangle^{72,A}$ in Table~\ref{tab1} are dominated by the uncertainties of the atomic factors. For comparison, the values of  $\delta\langle r^{2} \rangle^{72,A}$ based on the theoretically calculated atomic factors are also presented in Fig.~\ref{fig2}(a) (open
orange stars); they are within the systematic uncertainties of those obtained with calibrated atomic factors. The absolute charge radii of $^{68-74}$Ge listed in Table~\ref{tab1} are calculated using the charge radius of $^{72}$Ge from the electron scattering and muonic atom spectroscopy data~\cite{radiibook}.

\section{Discussion}
Figure~\ref{fig2}(a) presents a systematic comparison of the differential charge radii $\delta\langle r^{2} \rangle$ of several isotopic chains in this mass region. To better show nuclear structure effects we have subtracted  the spherical droplet model reference $\langle r^{2} \rangle_{0}$~\cite{Berdi1985}. Such obtained charge radii $\langle r^{2} \rangle-\langle r^{2} \rangle_{0}$ of nickel~\cite{Ni-radii2022}, copper~\cite{Cu-radii2020}, zinc~\cite{Zn-radii2019}, gallium~\cite{Ga-radii-2017} and germanium are shown in Fig.~\ref{fig2}(b) where odd and even-$N$ isotopes are represented by open and filled symbols, respectively. Although the charge radii of these five isotopic chains show a similar trend, appreciably larger variations in the charge radii and a significantly enhanced OES effect are seen for the germanium isotopes.

As discussed above, the light germanium isotopes around $N=40$ are soft systems in which shape coexistence phenomena are present. Deformation effects, whether related to static deformations or dynamic fluctuations, are expected to impact charge radii. To have a better understanding of the charge radii of $^{68-74}$Ge, we performed DFT calculations using three energy density functionals, namely two Fayans functional parametrizations Fy($\Delta r$, HFB)~\cite{K-radii2021,Fayan2017} and Fy(IVP)~\cite{In-radii}, as well as the Skyrme parametrization SV-min \cite{Ni-radii2022,SV-min}. All three functionals were optimized to the same large set of experimental data as described in Ref.~\cite{SV-min}. The functional Fy($\Delta r$, HFB) has been calibrated using the data on differential charge radii of calcium isotopes~\cite{Fayan2017}; this additional information substantially impacts the pairing functional and is crucial for reproducing the observed OES of charge radii. The trend of charge radii along the Cd isotopic chain was well reproduced~\cite{Hammen} with this functional. But in the mid-shell region of the semi-magic Sn and Ni isotopes, the slope of differential radii turned out to be overestimated~\cite{Sn-radii,Ni-radii2022} whereas the behavior at the shell closures was in good agreement with the experimental data~\cite{Sn-radii,Ni-radii}. This was attributed to missing isovector pairing correlations. In order to capture the underlying physics, the new functional Fy(IVP) was developed in Ref.~\cite{In-radii} by using the additional data on charge radii of calcium and lead isotopes.

\begin{figure}
\centerline{\includegraphics[width=\linewidth]{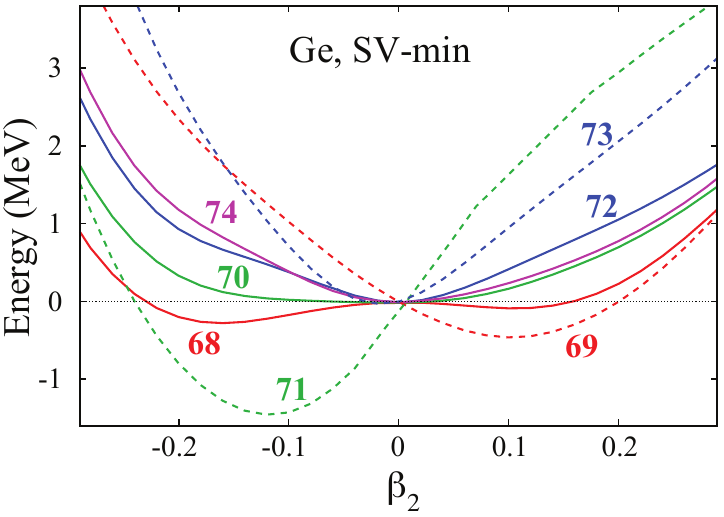}}
\vspace{-3mm}
\caption{
Potential energy computed with SV-min as function of
quadrupole deformation $\beta_2$ for $^{68,70,72,74}$Ge (full lines) and $^{69,71,73}$Ge (dashed lines). The energy is normalized to zero at the spherical shape.}
\label{fig:pes}
\vspace{-5mm}
\end{figure}

\begin{figure*}[t!]
\centering
\includegraphics[width=0.98\textwidth]{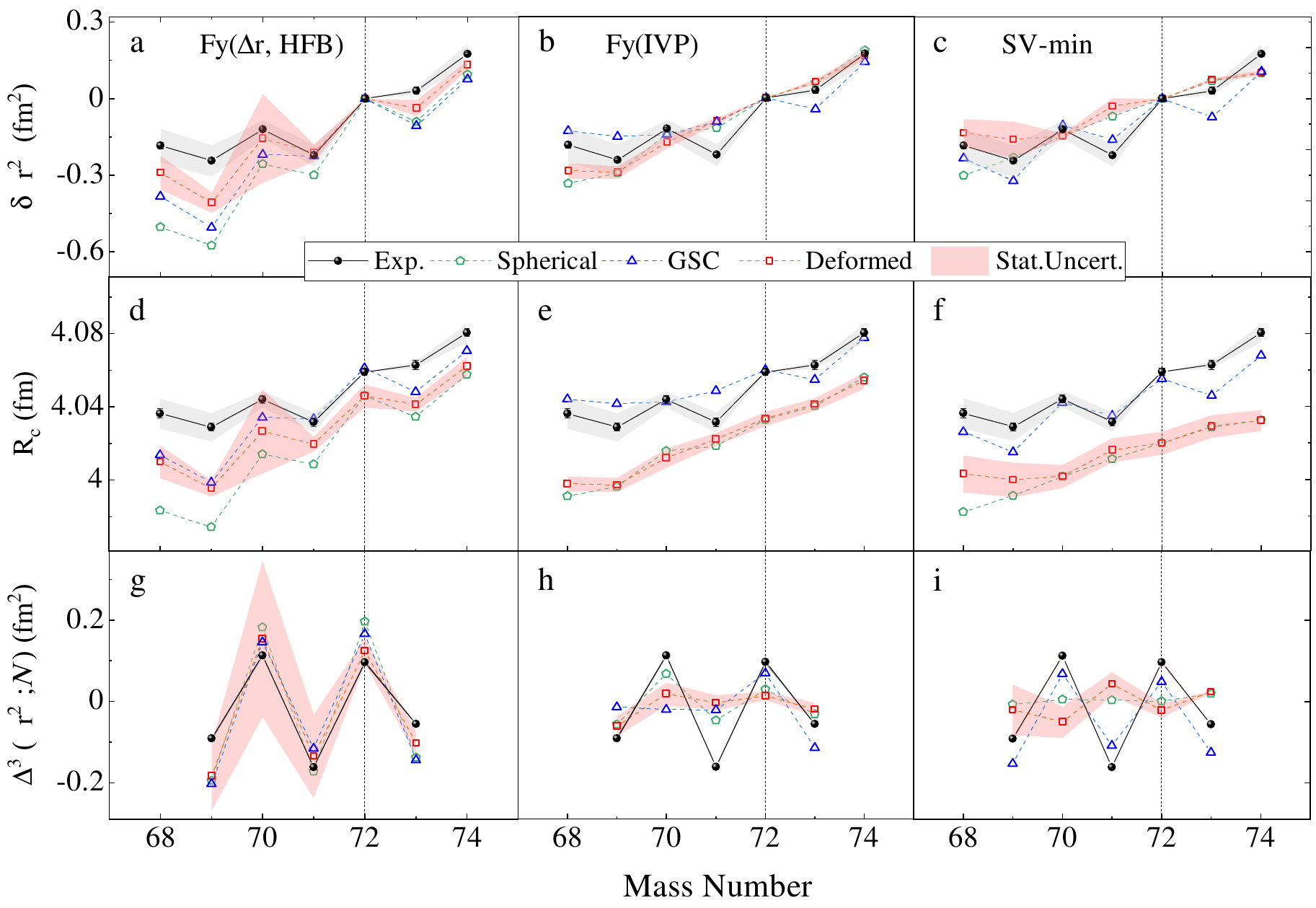}
\vspace{-3mm}
\caption{(a-c) Differential charge radii $\delta\langle r^{2} \rangle$,  (d-e) charge radii $R_{c}$ and (g-i) $\Delta^{3}(\langle r^{2}\rangle,N)$ of germanium isotopes compared to DFT results using Fy($\Delta r$, HFB), Fy(IVP), and SV-min energy density functionals. The statistical errors on the theoretical values come from the calibration uncertainties of the parametrizations. One can also consider the difference between Spherical and GSC variants as a rough estimate of systematic errors. See text for more details.}
\label{fig:dft}
\vspace{-1mm}
\end{figure*}

\begin{figure*}[t!]
\includegraphics[width=0.98\textwidth]{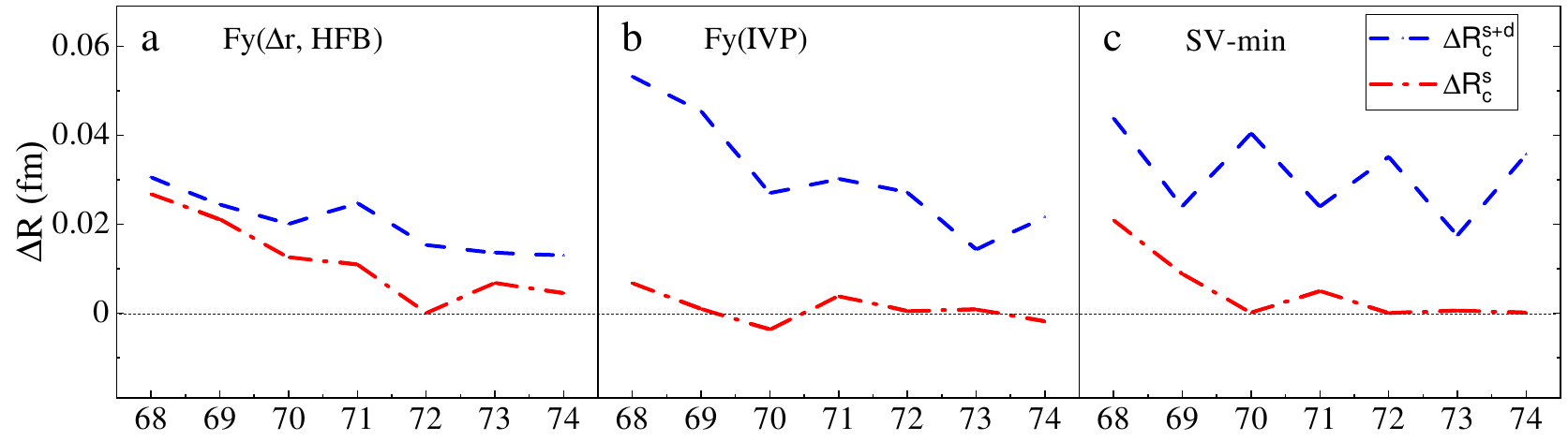}
\vspace{-3mm}
\caption{(a-c) Contribution of the static and dynamic deformation to the nuclear charge radii of $68-74$Ge isotopes calculated by Fy($\Delta r$, HFB), Fy(IVP), and SV-min functionals. See text for more details. }
\label{fig:fig5}
\vspace{-3mm}
\end{figure*}

As a representative example of DFT calculations for the germanium isotopes, Fig.~\ref{fig:pes} shows the potential-energy curves for $^{68-74}$Ge calculated with SV-min along the dimensionless quadrupole deformation $\beta_2$ deduced from the calculated intrinsic proton quadrupole moments:
\begin{equation}\label{eqbeta2}
\beta_2=4\pi\frac{\langle r^2Y_{20}\rangle}{3ZR^2} \;,\; R=1.2A^{1/3}\,\mathrm{fm}\;.
\end{equation}
The energy curves for even-even isotopes are extremely soft. As a consequence, their ground-state deformations are not well defined; the corresponding collective wave functions are in fact coherent superpositions of nearly-degenerate states having different quadrupole deformations. For the odd-$A$ isotopes, deformations are better defined due to a strong quadrupole polarization exerted by an odd neutron occupying a deformed single-particle level. In our calculations, we assumed that the quantum numbers $K^\pi$ of the blocked one-quasiparticle state correspond to the observed ground-state quantum numbers $I^\pi$ given in Table\,\ref{tab1}. As seen in Fig.~\ref{fig:pes}, the quadrupole polarization effects are substantial, and they strongly depend on the geometric character of the blocked orbital \cite{Nazarewicz1990,Schunck2010}.

It is thus not straightforward to unambiguously determine the ground state configuration from the calculated mean-field minima alone. In fact, the situation shown in Fig.~\ref{fig:pes} calls for a multi-reference theory such as the Generator Coordinate Method (GCM). Here, we handle GCM at the level of the Gaussian overlap approximation~\cite{Reinhard1987}. The actual procedure is explained in Refs.~\cite{Schmid1991,Kluepfel2008}. We compute the collective deformation path along axially symmetric shapes by quadrupole constrained DFT calculations
on a cylindrical grid in coordinate space~\cite{Reinhard2021}. This yields the collective potential energy curve as shown in Fig.~\ref{fig:pes} together with the corresponding collective mass and moment of inertia. The potential landscape is soft in triaxial
$\gamma$ direction which allows to interpolate potential energy, collective mass and inertia
into the full triaxial $\beta$-$\gamma$-plane. The ground-state wave function of the resulting collective Bohr Hamiltonian determines the GSC corrected radii. As a rule of thumb, the ground-state fluctuation of shapes cover typically a range of $\Delta\beta_2\approx 0.2$.
The final DFT results for the charge radii of germanium isotopes are presented in Fig.~\ref{fig:dft}. Results labeled ``Spherical'', ``Deformed'' and ``GSC'', represent calculations at spherical shape, minimum-energy configuration, and with GSC added, respectively. The statistical errors are computed based on the calibration uncertainties of the energy density functionals \cite{Dobaczewski2014,Erler2015}. As a rough estimate of systematic errors one can consider the difference between ``Spherical'' and ``GSC'' variants.

As seen in Fig.~\ref{fig:dft} (a-f), considering the theoretical and experimental uncertainties, all three functionals provide a fair agreement with experimental data for $\delta\langle r^{2} \rangle$ and $R_{\rm c}$. It is to be noted that, in general,
$R_{\rm c}^{\rm Spherical}<R_{\rm c}^{\rm Deformed} < R_{\rm c}^{\rm GSC}$. Indeed, since the
rms charge radius depends on quadrupole correlations as \mbox{$\langle Q^2\rangle \propto \langle \beta_2^2 \rangle$}, the difference
\mbox{$\Delta R_{\rm c}^{\rm s} \equiv R_{\rm c}^{\rm Deformed}-R_{\rm c}^{\rm Spherical}$} measures the contribution from the static deformation, while the \mbox{$\Delta R_{\rm c}^{\rm s+d} \equiv R_{\rm c}^{\rm GSC}-R_{\rm c}^{\rm Spherical}$} also measures the impact of dynamic fluctuations. As shown in Fig.~\ref{fig:fig5}, it is interesting to see that $\Delta R_{\rm c}^{\rm s+d}$ changes rather gradually for even-even isotopes, which is consistent with the data on quadrupole invariants $\langle Q^2 \rangle$
\cite{Sugawara2003,Ayangeakaa2016}. The relation between the three radii, $R_{\rm c}^{\rm Spherical}$, $R_{\rm c}^{\rm Deformed}$, and $R_{\rm c}^{\rm GSC}$ reflects the character of underlying potential energies. For Fy(IVP) and SV-min $\Delta R_{\rm c}^{\rm s}$ is smaller and $\Delta R_{\rm c}^{\rm s+d}$ is larger as $^{68-74}$Ge isotopes are predicted to be very soft in these models. On the other hand, for Fy($\Delta r$, HFB), $\Delta R_{\rm c}^{\rm s}$ and $\Delta R_{\rm c}^{\rm s+d}$ is similar as this model predicts smaller dynamic deformations.

While the performance of the three functionals for the general trend of $\delta\langle r^{2} \rangle$ and $R_{\rm c}$ is rather similar, the local variation of the nuclear charge radii, namely the OES defined as $\Delta^{3}(\langle r^{2}\rangle,N)=\langle r^{2}\rangle^{N} -\frac{1}{2}(\langle r^{2}\rangle^{N-1}+\langle r^{2}\rangle^{N+1})$, shows more model dependence (Fig.~\ref{fig:dft}(g-i)). Specifically, the relatively large OES observed in the germanium chain is captured by the Fy($\Delta r$, HFB) functional in all three calculation variants, and is very well described by the ``GSC'' variant of the SV-min model. Overall, the calculations that consider zero-point correlations describe the OES of the germanium radii fairly well. This indicates that the observed large OES in $^{68-74}$Ge can be attributed to both pairing and shape polarization effects, which -- as shown in Fig.~\ref{fig:pes} -- are appreciable in the transitional germanium isotopes.

\section{Summary and conclusion}
The charge radii of $^{68-74}$Ge isotopes were measured employing the collinear laser spectroscopy technique. By comparing with the neighbouring isotopic chains, we found that charge radii of germanium isotopes around $N=40$ exhibit large OES. Quantified DFT calculations using three different energy density functionals were augmented by considering ground-state quadrupole correlations. Within the uncertainties, all calculations are consistent with the experimental data on differential and absolute charge radii of $^{68-74}$Ge. The OES of the charge radii of germanium isotopes is better described by the calculation including the GSC. One can thus conclude that the behavior of charge radii along the Ge isotopic chain is strongly influenced by the polarization effects due to both pairing and shape deformation. This work also highlights the recent progress in the quantified DFT calculations for the deformed open shell nuclei.

\section*{Acknowledgments}
We acknowledge the support of the ISOLDE collaboration and technical teams. This work was supported by the National Key R\&D Program of China (Contract No. 2022YFA1604800, 2023YFA1606403), the National Natural Science Foundation of China (Nos:12027809,12350007); the BriX Research Program No. P7/12, FWO-Vlaanderen (Belgium), GOA 15/010 from KU Leuven; the UK Science and Technology Facilities Council grants ST/L005794/1 and ST/P004598/1; the BMBF Contracts No 05P18RDCIA, 05P21RDCI1, the Max-Planck Society, the Helmholtz International Center for FAIR (HIC for FAIR); the EU Horizon2020 research and innovation programme through ENSAR2 (No. 654002); the computing center RRZE of the Friedich-Alexander university Erlangen; and by the U.S. Department of Energy under Award Number DE-SC0013365 (Office of Science) and DE-SC0023175 (Office of Science, NUCLEI SciDAC-5 collaboration).

\bibliographystyle{elsarticle-num}
\bibliography{Ge-radii}

\end{document}